\documentclass[useAMS,usenatbib]{mn2e}
\usepackage{graphicx}

\newcommand{\Mjup}{\mbox{M$_{\rm Jup}$}}

\newcommand{\Msun}{\mbox{M$_{\odot}$}}

\newcommand{\ltsimeq}{\raisebox{-0.6ex}{$\,\stackrel
         {\raisebox{-.2ex}{$\textstyle <$}}{\sim}\,$}}
\newcommand{\gtsimeq}{\raisebox{-0.6ex}{$\,\stackrel
         {\raisebox{-.2ex}{$\textstyle >$}}{\sim}\,$}}
\newcommand{\nsvs}{NSVS\,1425}
\newcommand{\mnras}{MNRAS}
\newcommand{\apj}{ApJ}
\newcommand{\aap}{A\&A}
\newcommand{\nat}{Nature}
\newcommand{\apjl}{ApJL}
\newcommand{\aj}{AJ}
\newcommand{\icarus}{Icarus}

\title[]{On the dynamical stability of the proposed planetary system 
orbiting NSVS\,14256825 }

\author[R.A. Wittenmyer, J. Horner, and J.P. Marshall]{Robert 
A.~Wittenmyer$^{1}$\thanks{E-mail: rob@phys.unsw.edu.au (RW)}, 
J.~Horner$^{1}$ and J.P. Marshall$^{2}$\\
$^{1}$Department of Astrophysics and Optics, School of Physics, 
University of New South Wales, Sydney 2052, Australia\\
$^{2}$Departmento F\'isica Te\'orica, Universidad Aut\'onoma de Madrid, Cantoblanco, 28049, Madrid, Spain}
\begin{document}

\date{Accepted  Received ; in original form }

\pagerange{\pageref{firstpage}--\pageref{lastpage}} \pubyear{2013}

\maketitle

\label{firstpage}

\begin{abstract}

We present a detailed dynamical analysis of the orbital stability of the 
two circumbinary planets recently proposed to orbit the evolved 
eclipsing binary star system NSVS\,14256825.  As is the case for other 
recently proposed circumbinary planetary systems detected through the 
timing of mutual eclipses between the central binary stars, the proposed 
planets do not stand up to dynamical scrutiny.  The proposed orbits for 
the two planets are extremely unstable on timescales of less than a 
thousand years, regardless of the mutual inclination between the 
planetary orbits. \\

For the scenario where the planetary orbits are coplanar, a small region 
of moderate stability was observed, featuring orbits that were somewhat 
protected from destabilisation by the influence of mutual 2:1 
mean-motion resonance between the orbits of the planets.  Even in this 
stable region, however, the systems tested typically only survived on 
timescales of order 1 million years, far shorter than the age of the 
system. \\

Our results suggest that, if there are planets in the NSVS\,14256825 
system, they must move on orbits dramatically different to those 
proposed in the discovery work.  More observations are clearly 
critically required in order to constrain the nature of the suggested 
orbital bodies.

\end{abstract}

\begin{keywords}

binaries: close, binaries: eclipsing, stars: individual: NSVS14256825, 
planetary systems, planets and satellites: dynamical evolution and stability

\end{keywords}

\section{Introduction}

The intriguing possibility of circumbinary planets has been a feature of 
science-fiction for decades.  Theoretical studies of planetary formation 
have shown that such systems are indeed possible \citep{ql06, gong13}, 
and that they can be dynamically stable \citep{kubala93, hw99}, but 
their secure detection remained quite elusive, even in this present age 
of accelerating exoplanetary discovery.  Finally, in late 2011, the 
first unambiguous example of a circumbinary planet was announced by the 
\textit{Kepler} science team \citep{doyle11}.  This discovery was 
quickly followed by the first multiple-planet circumbinary system, 
Kepler-47 \citep{jerry12a}, and two additional circumbinary systems 
\citep{jerry12b, welsh12}.  These exciting results have yielded a first 
estimate that $\sim$1\% of close binary stars host such circumbinary 
planets.

In addition to the \textit{Kepler} discoveries, there is a growing body 
of literature claiming the detection of circumbinary planets orbiting 
post-common-envelope host stars.  Such objects can, in theory, be 
detected by light-travel-time variations: as the postulated planetary 
companions orbit the central stars, they cause those stars to move back 
and forth as they orbit around the system's centre of mass.  As a 
result, the distance between the Earth and the host stars varies as a 
function of time, meaning that the light from the stars must sometimes 
travel further to reach us than at other times.  This effect results in 
measurable variations in the timing of mutual eclipse events between the 
two stars that can be measured from the Earth.  The stars proposed as 
hosts of circumbinary planets detected in this manner include 
cataclysmic variables (e.g. UZ\,For, Potter et al.~2011), 
pre-cataclysmic variables (NN\,Ser, Beuermann et al.~2010), and detached 
subdwarf-M dwarf (e.g. HW\,Vir, Beuermann et al.~2012) or subdwarf-white 
dwarf (e.g. RR\,Cae, Qian et al.~2012) binaries.  \citet{zoro13} provide 
a complete summary of the 12 proposed post-common-envelope planetary 
systems in their Table 2.

While the profusion of circumbinary planets claimed to orbit 
post-common-envelope binaries would suggest that such systems are 
extremely common, all is not as it seems.  When subjected to rigorous 
dynamical testing, a great many of these proposed planetary systems have 
been found to be simply unfeasible.  That is, the orbits of the proposed 
planets are such that, if they were truly planets, they would rapidly 
experience significant and ultimately catastrophic mutual interactions 
(ejections or collisions).  Recently, several of the proposed 
circumbinary multiple-planet systems have been tested in this manner.  
Highly detailed dynamical maps, showing the lifetimes of a wide range of 
orbital configurations for the planets in question, demonstrate that a 
number of these candidate systems are dynamically unstable on timescales 
of $\ltsimeq 10^4$ yr.  Some proposed circumbinary systems investigated 
in this way include HU Aqr \citep{horner11, HUAqr, hinse12}, NN Ser 
\citep{NNSer}, and HW Vir \citep{HWVir}.  The same dynamical mapping 
techniques have also been applied to multiple-planet systems discovered 
by radial velocity studies, in order to check their long-term stability 
\citep{142paper} and to assess the role of low-order resonances 
\citep{texas1, texas2, subgiants}.

NSVS\,14256825 (hereafter \nsvs) is an eclipsing binary consisting of a 
subdwarf OB star and an M dwarf orbiting one another with a period of 
0.110374~days \citep{almeida12}.  The combination of a hot subdwarf and 
a late-type dwarf produces significant reflection effects in the light 
curves, and such systems are known as ``HW Vir'' systems, of which only 
10 are currently known \citep{nsvspaper}.  Since its discovery in the 
Northern Sky Variability Survey \citep{woz04}, eclipse timings have been 
reported by \citet{wils07}, \citet{kil12}, and \citet{b12a}.  A linear 
period change was noted by \citet{kil12}, and \citet{b12a} reported a 
cyclic period change, which they attributed to the presence of a 
$\sim$~12~\Mjup\ planet with an unconstrained period $P\gtsimeq$20 yr.  
Most recently, \citet{nsvspaper} presented additional eclipse timings 
and, by combining all available timings, fitted two periodicities which 
were then attributed to the influence of two unseen circumbinary 
planets.  The reported parameters for the candidate planets are given in 
Table~\ref{params}.  In this work, we bring our well-tested dynamical 
mapping techniques (Section 2) to bear on the candidate \nsvs\ planetary 
system.  We determine the dynamical stability of the complete $\pm 
3\sigma$ range of orbital parameters, and we test the effect of mutual 
inclinations between the two planets (Section 3).  We discuss the 
results and make our conclusions in Section 4.

\begin{table}
  \centering
  \caption{Data on the \nsvs\ system from \citet{nsvspaper}.}
  \begin{tabular}{lll}
  \hline
Parameter & Inner Planet & Outer Planet \\
 \hline
Eccentricity                     & 0.0$\pm$0.08  & 0.52$\pm$0.06 \\
$\omega$ (degrees)               & 11.4$\pm$7    & 97.5$\pm$8 \\
Orbital Period (yrs)             & 3.49$\pm$0.21 & 6.86$\pm$0.25 \\
Orbital Radius (AU)              & 1.9$\pm$0.3   & 2.9$\pm$0.6 \\
$M\sin i$ (\Mjup)                & 2.8$\pm$0.3   & 8.0$\pm$0.8 \\
 \hline
 \end{tabular}
\label{params}
\end{table}

\section{Dynamical Analysis}

As in our previous work \citep[e.g. ][]{marshall10,horner11,NNSer}, we 
used the Hybrid integrator within the \textit{N}-body dynamics package 
{\sc Mercury} \citep{chambers99} to perform our integrations.  We held 
the initial orbit of the inner planet fixed at its best-fit parameters, 
as reported in \citet{nsvspaper}, and then created 126,075 test systems. 
In those test systems, the initial orbit of the outer planet was varied 
systematically in semi-major axis $a$, eccentricity $e$, periastron 
argument $\omega$, and mean anomaly $M$, resulting in a $41x41x15x5$ 
grid of ``clones'' spaced evenly across the 3$\sigma$ range in those 
parameters (as given in Table~\ref{params}).  For the mean anomaly, not 
reported in \citet{nsvspaper}, we simply tested the entire allowed range 
0-360 degrees.  We assumed the central stars to be a single point mass, 
an acceptable approximation as the binary separation is $\ll$ orbital 
radius of the inner planet, with $M=0.62$\Msun\ \citep{b12a}\footnote{We 
note that \citet{almeida12} recently determined a combined mass of 
0.528$\pm$0.074 for the central stars, $1.2\sigma$ smaller than our 
assumed value. This small difference would have no effect on the 
interactions \textit{between} the planets, only on the overall 
\textit{scale} of the system, and hence does not affect our results.}.  
We assumed the planets were coplanar with each other and had masses 
equivalent to their minimum mass, $M\sin i$.  We then followed the 
dynamical evolution of each test system for a period of 100 million 
years, and recorded the times at which either of the planets was removed 
from the system.  Planets were removed if they collided with one 
another, hit the central body, or reached a barycentric distance of 
10~AU.

In addition to this highly detailed ``nominal'' run, we examined the 
effects of mutual inclinations between the two planets by running 
further suites of simulations at lower resolution \citep[e.g. 
][]{horner11, subgiants}.  We varied the initial orbit of the outer 
planet as above, resulting in a $21x21x5x5$ grid in $a$, $e$, $\omega$, 
and $M$, respectively (for a total of 11,025 test systems).  We ran five 
such scenarios, with the inclination between the two planets set at 5, 
15, 45, 135, and 180 degrees (the latter two corresponding to a 
retrograde orbit for the outer planet).  Again, the 11,025 test systems 
were allowed to run for 100 million years, or until the system 
was destabilised due to ejection or collision.

\section{Results}

\begin{figure}
\includegraphics[scale=0.36]{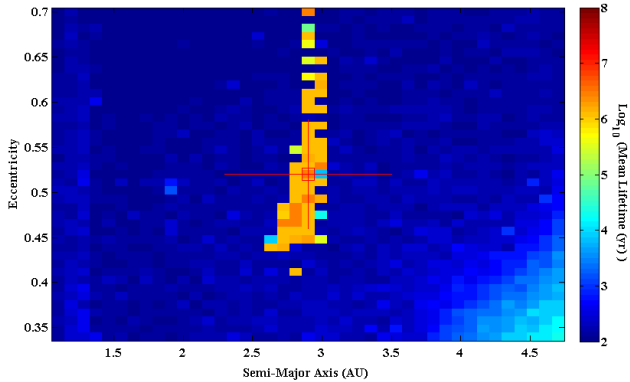}
\caption{Dynamical stability of the \nsvs\ system as proposed by 
\citet{nsvspaper}, as a function of the semi-major axis, $a$, and 
eccentricity, $e$, of the outer planet.  The mean lifetime of the 
planetary system (in $log_{10}$ (lifetime/yr)) at a given $a-e$ 
co-ordinate is denoted by the colour of the plot.  The lifetime at each 
$a-e$ location is the mean value of 75 separate integrations carried out 
on orbits at that $a-e$ position (testing a combination of 15 unique 
$\omega$ values, and 5 unique $M$ values).  The nominal best-fit orbit 
for the outer planet is shown as the small red square with 
$\pm~1$-$\sigma$ error bars.}
\label{nominal}
\end{figure}

Figure~\ref{nominal} shows the stability of the nominal orbits for the 
\nsvs\ system as given in \citet{nsvspaper}.  It is immedately apparent 
that the vast majority of the $\pm 3\sigma$ parameter space is extremely 
unstable, with mean lifetimes of less than 1000 yrs.  This 
result is not particularly surprising, given that the great majority of 
solutions tested place the two planets on mutually crossing orbits - 
meaning that close encounters between the two are a certainty.

The nominal best-fit orbit of the outer planet does fall in a narrow 
strip of increased stability, featuring mean lifetimes $\sim 10^6$yr.  
However, as the subdwarf B host star has evolved well off the main 
sequence, one would expect the system to be considerably older than this 
mean lifetime.  While the ages of subdwarf B stars are highly uncertain 
\citep{hu10}, even an A-type progenitor would make the system $\sim$0.5 
Gyr old.  Population-synthesis models of the close binary progenitors of 
these post-common-envelope systems show that the initial mass 
distribution of primary stars that result in subdwarf B stars ranges 
from 1.0-1.8 \Msun.  The shortest-lived (i.e. highest-mass) progenitors 
would then evolve off the main sequence in $\sim 2\times\,10^9$ yr.  
Hence, any planets which existed before the common-envelope stage would 
be expected to demonstrate dynamical stability on a timescale far longer 
than that exhibited by the \nsvs\ system.

The observed strip of moderate stability is attributed to the protective 
influence of the mutual 2:1 mean-motion resonance between the two 
proposed planets.  Such resonant protection is well known, both in our 
own Solar system \citep[e.g. ][]{QR322,TrojanII}, and in exoplanetary 
science \citep[e.g. ][]{laplace,texas1,texas2}.  Indeed, it is 
interesting to compare our results with those obtained for the proposed 
planets around HU~Aqr \citep[e.g. Figure 1 of ][]{horner11}.  In both 
cases, the proposed outer planet lies on a highly eccentric best-fit 
orbit ($e \sim 0.5$) that crosses that of the innermost planet and the 
two planets are close to mutual 2:1 mean-motion resonance. Additionally, 
the proposed system proves to be dynamically unstable on typical 
timescales of hundreds or thousands of years.

\begin{figure*}
\includegraphics[scale=0.82]{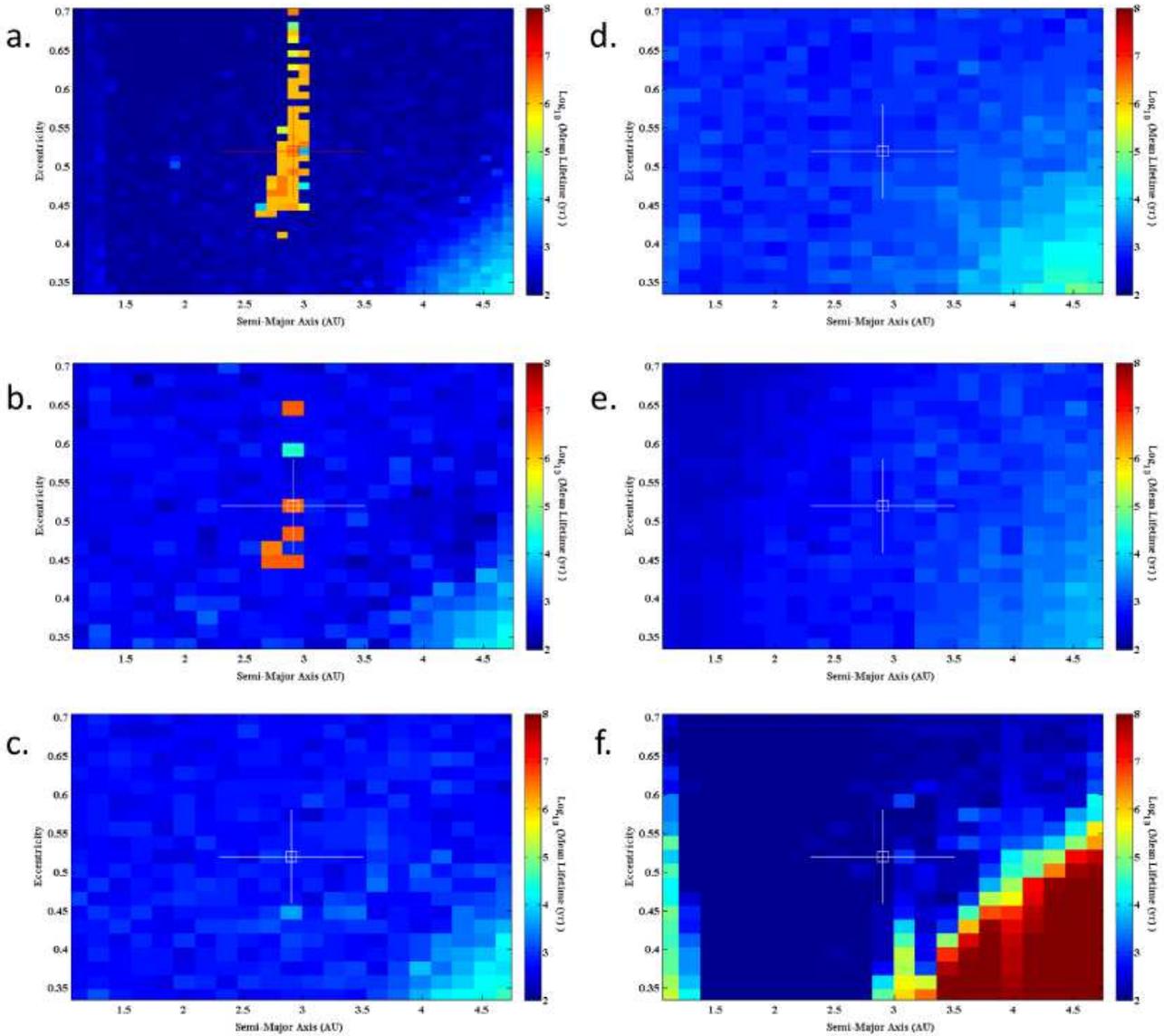}
\caption{Dynamical stability for the \nsvs\ system, for six values 
of inclination between the two planets.  Panels (a) through (f) 
represent mutual inclinations of 0, 5, 15, 45, 135, and 180 degrees, 
respectively.  Panel (a) is a duplicate of Figure~\ref{nominal}, 
reprised here for ease of comparison.  As in the previous figure, the 
colour bar represents the log of the mean survival time at each $a-e$ 
position (testing a combination of 5 $\omega$ values and 5 $M$ values).}
\label{inclined}
\end{figure*}

We then considered the possibility of non-zero mutual inclinations 
between the two candidate planets in the \nsvs\ system.  For a 
significantly interacting planetary system, retrograde orbits can 
provide a dynamically stable configuration \citep[e.g. ][]{eberle10, 
horner11, q12, morais12}.  To explore the dynamical stability of systems 
resulting from such non-coplanar scenarios, we performed a second suite 
of simulations at lower resolution, as described in Section 2.  The 
results of the five mutually-inclined scenarios are shown in 
Figure~\ref{inclined}, along with the nominal coplanar scenario from 
Figure~\ref{nominal}.  We see that mutually-inclined scenarios are even 
more unstable than the coplanar scenario.  Even for the 180-degree 
configuration (i.e. retrograde and coplanar, panel f), the only region 
of stability appears in the lower right, at the lowest eccentricities 
and largest semi-major axis for the outer planet.  As was the case for 
HU Aquarii \citep{horner11}, the highly stable region is restricted to 
orbits which have periastron distances several Hill radii from the inner 
planet.  However, this region lies well beyond the 1$\sigma$ 
uncertainties of the orbits derived by \citet{nsvspaper}.

\section{Discussion and Conclusions}

We have shown that the two circumbinary planets proposed by 
\citet{nsvspaper} are dynamically unfeasible in virtually any 
configuration within $3\sigma$ of their derived orbital parameters.  
Based on our results, a mechanism other than, or in addition 
to, one (or more) planets is needed to explain the observed period 
variations.  \citet{zoro13} note that about 90\% of eclipsing 
post-common-envelope binaries are reported to have timing variations, 
nearly all of which have been attributed to one or more orbiting 
circumbinary planets.  Both avenues to such planetary systems -- 
survival of the planets through the Asymptotic Giant Branch (AGB) and 
planetary nebula phase, or secondary accretion from the ejected envelope 
of the primary star -- are possible but highly uncertain \citep[e.g. 
][]{postnov92, tavani92, phinney93, vl07, hansen09, veras11, kunitomo11, 
mv12}.

\citet{zoro13} examined the formation and observational statistics of 
circumbinary planets in order to investigate the puzzling trend that 
fully 90\% of post-common-envelope eclipsing binaries appear to host 
planets.  They note that observations of disk-bearing pre-main-sequence 
stars \citep{kraus12} indicate that the circumbinary disk lifetime is 
too short ($\ltsimeq$1 Myr) to form giant planets by core accretion.  
Observational results from \textit{Kepler} also point to a main-sequence 
circumbinary giant planet frequency of $\sim$1\% \citep{welsh12}.  
\citet{jupiters} estimate the frequency of giant planets in 3-6 AU 
orbits to be no more than 37\%, based on data on single main-sequence 
stars from the Anglo-Australian Planet Search.  All of these strands of 
evidence led \citet{zoro13} to the conclusion that ``virtually all 
close-compact binaries...are unlikely to be explained by 
first-generation planets.''

Another potential solution to the near-ubiquitous presence of planets 
around evolved binaries is the formation of these companions in the post 
main sequence circumstellar envelope produced by the subdwarf OB 
progenitor as it evolves through its AGB phase.  The amount of material 
cast off by an AGB star \citep[up to 70\% of the stellar 
mass,][]{agb_ml} is comparable to that of the Minimum Mass Solar Nebula 
\citep{mmsn} and the lifetime of the AGB phase is similar to that of 
gas-rich protoplanetary discs within which first generation planet 
formation occurs \citep{h07}, lending plausibility to the idea of a 
second generation of planet formation.  However, this second generation 
planet formation scenario is highly speculative and requires further 
scrutiny (and detailed modelling) before it can be considered a viable 
answer \citep{as08,p10}.

As was the case in earlier studies of proposed circumbinary planets in 
highly evolved systems \citep[e.g. ][]{horner11,HUAqr,HWVir}, we find 
that the planetary system proposed in the evolved binary system 
NSVS\,14256825 is simply not dynamically feasible.  Our results suggest 
that some mechanism other than planets must be responsible for the 
observed eclipse-timing variations, and once again highlight the 
critical importance of performing detailed dynamical analyses of 
potential new planetary systems in order to determine whether the 
proposed systems make sense.

\section*{Acknowledgments}

JPM is supported by Spanish grant AYA 2011/02622.  The work was 
supported by iVEC through the use of advanced computing resources 
located at the Murdoch University, in Western Australia. This research 
has made use of NASA's Astrophysics Data System (ADS), and the SIMBAD 
database, operated at CDS, Strasbourg, France.


\label{lastpage}

\end{document}